# First-principles study of superconducting ScRhP and ScIrP pnictides

M. T. Nasir[a], M. A. Hadi[b*], M. A. Rayhan[a], M. A. Ali[c], M. M. Hossain[c], M. Roknuzzaman[d], S. H. Naqib[b*], A. K. M. A. Islam[b*,e], M. M. Uddin[c], K. Ostrikov[d]

[a]Department of Arts & Science, Bangladesh Army University of Science and Technology, Saidpur-5310, Nilphamari, Bangladesh.
[b]Department of Physics, University of Rajshahi, Rajshahi-6205, Bangladesh.
[c]Department of Physics, Chittagong University of Engineering and Technology, Chittagong-4349, Bangladesh.
[d]School of Chemistry, Physics and Mechanical Engineering, Queensland University of Technology, QLD 4000, Australia
[e]International Islamic University Chittagong, 154/A College Road, Chittagong, Bangladesh

**Abstract**

For the first time, we have reported in this study an *ab initio* investigation on elastic properties, Debye temperature, Mulliken population, Vickers hardness, and charge density of superconducting ScRhP and ScIrP phosphides. The optimized cell parameters show fair agreement with the experimental results. The elastic constants and moduli, Poisson's as well as Pugh's ratio and elastic anisotropy factors have also been calculated to understand the mechanical behaviors of these ternary compounds. Their mechanical stability is confirmed via the calculated elastic constants. The calculated values for Poisson's and Pugh's ratio indicate the ductile nature of these compounds. ScIrP is expected to be elastically more anisotropic than ScRhP. The estimated value of Debye temperature predicts that ScRhP is thermally more conductive than ScIrP and the phonon frequency in ScRhP is higher than that in ScIrP. The hardness of ScRhP is lower due to the presence of antibonding Rh-Rh in ScRhP. The investigated electronic structures predict that the metallic conductivity of ScRhP reduces significantly when Rh is replaced with Ir. The main contribution to the total density of states (TDOS) at Fermi-level ($E_F$) comes from the *d*-electrons of Sc and Rh/Ir in both compounds. These two ternary compounds are characterized mainly by metallic and covalent bonding with little ionic contribution. As far as superconductivity is concerned, the matrix elements of electron-phonon interaction are noticeably enhanced in ScIrP compared to that in ScRhP.

## 1. Introduction

There are many ternary pnictides that crystallize in the ordered hexagonal Fe$_2$P-type structure with a chemical formula of $MM'X$ ($X$ = P and As) and a space group of $P\bar{6}2m$ [1]. In their chemical formula, $M$ stands for early transition metals including Ca and $M'$ is commonly a late transition metal.

A large number of pnictides with this structure exhibit superconducting behaviors with relatively high transition temperature $T_c$. The most well known members in this family ZrRuP, ZrRuAs and HfRuP exhibit superconductivity with an onset transition temperature $T_c \sim$ 12 K [2–7]. Another member in this family, MoNiP is observed to show a bulk superconducting transition at 15.5 K [8,9]. In fact, the superconductivity in $MM'X$ raised the attention of the scientific community to this family. Very recently, Okamoto *et al*. [10] synthesized the ternary phosphide ScIrP and confirmed its superconducting transition temperature of 3.4 K. Following this study, Inohara *et al*. reported the discovery of ScRhP, which shows superconductivity with $T_c \sim$ of 2 K [11]. They also discuss the distinguishing features of superconducting state in ScRhP by means of comparison with its isoelectronic and isostructural ScIrP. The comparison shows that $T_c$ of ScRhP is nearly half that of ScIrP, whereas the upper critical field $H_{c2}(0)$ of ScRhP is found to be only ~1/10 of that ScIrP. The low $H_{c2}(0)$ of ScRhP indicates the weaker spin–orbit interaction of the Rh 4d electrons in ScRhP compared to the Ir 5d electrons in ScIrP. In addition, the upper critical field $H_{c2}(0)$ in ScIrP is considerably increased by the antisymmetric spin–orbit interaction of the Ir 5d electrons in the noncentrosymmetric crystal structure. Additionally, the electron–phonon couplings in both ternary phosphides are recommended to be weak or moderate from the fact that the experimentally measured Sommerfeld coefficients, *i.e.,* $\gamma_{expt}$ values are nearly the same as $\gamma_{cal}$ calculated using the first principles methods. If the conventional phonon-mediated superconductivity is understood in both phosphides, the larger density of states $N(E_F)$, the higher phonon frequency $\omega_p$, and the stronger electron–phonon interaction will lead to a higher $T_c$. Except superconducting properties, there is hardly any study on other physical properties of these phosphide superconductors. The elastic properties, Mulliken population, Vickers hardness of these ternaries are still unexplored. Only band structure and density of states (DOS) among electronic properties are calculated for both the compounds [10–12].

---

[*]Corresponding author: email: hadipab@gmail.com, azi46@ru.ac.bd, salehnaqib@yahoo.com

So, here we plan to conduct first-principles study of various physical properties mentioned above of these ternary phosphide pnictides. The previous study [11] shows that the DOSs at the Fermi level calculated with and without spin-orbit coupling (SOC) are almost same for both superconductors ScRhP (9.58 and 9.61 states per eV, respectively) and ScIrP (5.16 and 4.99 states per eV, respectively). Moreover, the inclusion of SOC has only a minor effect on structural, elastic and bonding properties of transition metal based compounds such as MAX phases, namely, $M_2AlC$ (M = Ti, V, and Cr) and $Mo_2AC$ (A = Al, Si, P, Ga, Ge, As, and In) [13,14]. For these reasons, SOC has not been taken into consideration in the present study.

## 2. Computational Methods

The calculations are carried out using the Cambridge Serial Total Energy Package (CASTEP) code [15] based on the first-principles density-functional theory (DFT) [16]. The generalized gradient approximation (GGA) of Perdew-Burke-Ernzerhof (PBE) [17] is used to evaluate the electronic exchange and correlation potentials. The electrostatic interaction between valence electron and ionic core is represented by the Vanderbilt-type ultrasoft pseudopotentials [18]. The cutoff energy for the plane wave expansion is chosen as 440 eV. A k-point mesh of $8 \times 8 \times 12$, according to Monkhorst-Pack [19] scheme, is used for integration over the first Brillouin zone. The Broyden–Fletcher–Goldfarb–Shanno (BFGS) algorithm [20] is applied to optimize the atomic configuration and density mixing is used to optimize the electronic structure. Convergence tolerance for energy, maximum force, maximum displacement, and maximum stress are chosen as $5.0 \times 10^{-6}$ eV/atom, 0.01 eV/Å, $5.0 \times 10^{-4}$ Å, and 0.02 GPa, respectively.

## 3. Results and Discussion

### 3.1. *Structural properties*

The ternary phosphides ScRhP and ScIrP crystallize in a hexagonal structure with space group of $P\bar{6}2m$. (No 187). The structures are fully relaxed by optimizing the geometry with the lattice parameters and internal coordinates. In optimized structure, the Sc atom resides on the 3g Wyckoff position with fractional coordinates (0.5831, 0, 0.5) and (0.5805, 0, 0.5) in ScRhP and ScIrP, respectively. In ScRhP and ScIrP, the Rh and Ir atoms occupy 3f Wyckoff site with fractional coordinates (0.2508, 0, 0) and (0.2512, 0, 0), respectively. The P atoms occupy two Wyckoff positions 1b and 2c with fractional coordinates (0, 0, 0.5) and (1/3, 2/3, 0), respectively in both compounds. The unit cell of ScRhP as a structural model of hexagonal *MM′X* crystals is shown in Fig. 1 and the unit cell properties are given in Table 1. The calculated lattice constants *a* and *c* and unit cell volume *V* are found to be consistent with the experimental values [10,11,21]. It is observed that the replacement of Rh by Ir atom affects the lattice constants; the unit cell volume remains almost unchanged. The lattice constant *a* is found to decrease by 2.14%, whereas the lattice parameter *c* increases by 3.27% when Rh is substituted by Ir atom. Consequently, *c/a* ratio increases in ScIrP by 5.47% compared to that in ScRhP, which is indicative of highly compressive in crystal structure of ScRhP in the *c*-direction in comparison with that in ScIrP.

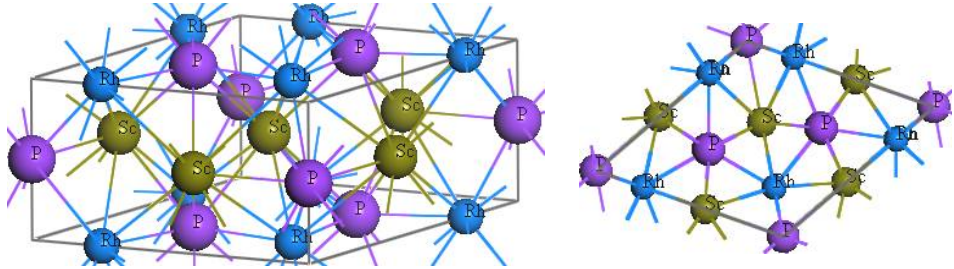

**Fig. 1**: Optimized unit cell of ScRhP (left) with its two dimensional view in *ab* plane (right). ScIrP is iso-structural with Ir atoms on place of Rh.

**Table 1.** Optimized lattice parameters (*a* and *c*, in Å), hexagonal ratio *c/a*, and unit cell volume (*V* in Å$^3$) of ScRhP and ScIrP along with experimental values.

| Phase | *a* | *c* | *c/a* | *V* | Remarks |
|---|---|---|---|---|---|
| ScRhP | 6.481 | 3.790 | 0.585 | 137.857 | This calc. |
|  | 6.453 | 3.727 | 0.578 | 134.404 | Expt. [11] |
| ScIrP | 6.342 | 3.914 | 0.617 | 136.326 | This calc. |
|  | 6.331 | 3.885 | 0.614 | 134.855 | Expt. [10] |
|  | 6.372 | 3.892 | 0.611 | 136.853 | Expt. [21] |

3.2. *Mechanical properties*

To be mechanically stable, the hexagonal crystals should fulfill the Born criteria [22]: $C_{11} > 0$, $C_{11} - C_{12} > 0$, $C_{44} > 0$, $(C_{11} + C_{12})C_{33} - 2C_{13}C_{13} > 0$. The calculated five independent elastic tensors, shown in Table 2, completely satisfy the above conditions, which indicate that the hexagonal ScRhP and ScIrP are mechanically stable. Unfortunately, there is no experimental data for elastic constants to compare with. It is observed that the unidirectional elastic tensors $C_{11}$ and $C_{33}$ are higher than the pure shear elastic constant $C_{44}$ for both ternaries. It means that the shear deformation is easier than the linear compression along the crystallographic *a*- and *c*-axes. Again, $C_{11}$ is greater than $C_{33}$, which implies that the two ternary pnictides are more incompressible along *a*-axis compared to that along *c*-axis. The substitution of Rh by Ir causes a significant increase of $C_{11}$ and $C_{33}$ and reduction of $C_{44}$. The comparatively large value of $C_{44}$ for ScRhP indicates that the ability of resisting the shear deformation in (100) plane is significant in ScRhP compared to ScIrP. The elastic tensor $C_{12}$ together with $C_{13}$ combines a functional stress component in the crystallographic *a*-direction in the presence of a uniaxial strain along the crystallographic *b*- and *c*-axes, respectively. The reasonable values of these tensors imply that the ternary phosphides ScRhP and ScIrP are capable of resisting the shear deformation along the crystallographic *b*- and *c*-axes, while a large force is applied to the crystallographic *a*-axis. Though the elastic constant $C_{12}$ increases, the elastic tensor $C_{13}$ decreases when Rh is replaced with Ir.

The polycrystalline elastic properties, namely bulk modulus and shear modulus are calculated using single crystal elastic constants, $C_{ij}$ in the Voigt-Reuss-Hill approximations [23–25] and listed in Table 2. So, the compound ScIrP, with a higher *G*, should be more rigid compared to ScRhP. It is evident from Table 2 that both the novel phosphide superconductors should behave as ductile materials as their Pugh's ratios, $B/G > 1.75$ [26]. The comparison of the values of $Y$ (= $9BG/(3B + G)$) given in Table 2 shows that substitution of Rh with Ir increases the stiffness. In fact, the replacement of Rh with Ir increases all the moduli ($B$, $G$, and $Y$) of ScIrP including Pugh's ratio. Thus, the mechanical properties are enhanced significantly when Rh is substituted with Ir. The thermal shock resistance, which varies inversely with $Y$ [27] is an essential factor for selecting a material as a thermal barrier coating (TBC) substance. We observe that as $Y$ increases due to substitution of Ir, $R$ decreases considerably in ScIrP. Accordingly, ScRhP should be more resistant to thermal shock than ScIrP.

**Table 2**. The single crystal elastic constants ($C_{ij}$ in GPa), polycrystalline bulk, shear, and Young's modulus, ($B$, $G$ and $Y$ in GPa), Pugh's ratio $G/B$, Poisson's ratio $v$ and elastic anisotropy factors ($A_1$, $A_2$, $A_3$, $A_B$ and $A_G$) of ScRhP and ScIrP.

| Compound | Single crystal elastic constants | | | | |
|---|---|---|---|---|---|
| | $C_{11}$ | $C_{12}$ | $C_{13}$ | $C_{33}$ | $C_{44}$ |
| ScRhP | 277 | 105 | 138 | 228 | 87 |
| ScIrP | 351 | 124 | 115 | 303 | 53 |
| Compound | Polycrystalline bulk elastic properties | | | | |
| | $B$ | $G$ | $Y$ | $B/G$ | $v$ |
| ScRhP | 171 | 76 | 199 | 2.25 | 0.30 |
| ScIrP | 190 | 82 | 215 | 2.32 | 0.31 |
| Compound | Elastic anisotropy factors | | | | |
| | $A_1$ | $A_2$ | $A_3$ | $A_B$ | $A_G$ |
| ScRhP | 0.55 | 1.01 | 0.55 | 0.05 | 2.9 |
| ScIrP | 1.95 | 0.47 | 0.91 | 0.31 | 6.4 |

Poisson's ratio $v$ is calculated via the equation, $v = (3B - 2G)/(6B + 2G)$ and presented in Table 2. Frantsevich *et al*. [28] predicted that a material behaves in brittle manner if its Poisson's ratio does not exceed the value of 0.26 and if exceeds this critical value the material will show ductility. This criterion indicates that both the ternary pnictides are ductile in nature.

To identify the interatomic forces that stabilize the crystal systems, Poisson's ratio serves as a good predictor. There are two types of interatomic forces for stabilizing the crystal solids. Firstly, the central forces among the nearest neighbors are responsible for the stability of fcc and almost all bcc crystals. Secondly, the stability of the diamond structures assumes the non-central forces [29]. The structural stability will be established with central forces if the crystals have Poisson's ratio ranging from 0.25 to 0.50. The non-central forces will be active to stabilize the crystal structure when the systems have Poisson's ratio either less than 0.25 or greater than 0.50 [30]. The calculated value of $v$ for ScRhP and ScIrP lies within 0.25 and 0.50, indicating that the interatomic forces in the two ternary phosphides are basically central forces.

There are three types of shear anisotropy factors for hexagonal crystals due to having three independent shear elastic constants. These factors for {100}, {010} and {001} shear planes can be defined successively as [31]:

$$A_1 = \frac{(C_{11} + C_{12} + 2C_{33} - 4C_{13})}{6C_{44}}$$

$$A_2 = \frac{2C_{44}}{C_{11} - C_{12}}$$

$$A_3 = \frac{(C_{11} + C_{12} + 2C_{33} - 4C_{13})}{3(C_{11} - C_{12})}$$

The calculated values of these parameters listed in Table 2 signify that the studied compounds ScRhP and ScIrP are elastically anisotropic. The amount of deviation from unity (isotropic) indicates the degree of elastic anisotropy.

The percentage anisotropy in compressibility and shear in polycrystalline aggregates are calculated with the following equations [32]:

$$A_B = \frac{B_V - B_R}{B_V + B_R} \times 100\%$$

and

$$A_G = \frac{G_V - G_R}{G_V + G_R} \times 100\%$$

Here, $B$ and $G$ stand for the bulk and shear moduli and their subscripts $V$ and $R$ indicate the Voigt and Reuss limits, respectively. It is obvious from Table 2 that the anisotropy in shear is prominent than that in compressibility for both pnictides. For hexagonal crystals, another elastic anisotropy factor is important and which is derived from the ratio of linear compressibility coefficient along the $c$-axis to that along the $a$-axis: $k_c/k_a = (C_{11} + C_{12} - 2C_{13})/(C_{33} - C_{13})$. The calculated values of 1.18 and 1.30 for ScRhP and ScIrP indicate that the compressing along $c$-axis is easier than that along $a$-axis. The effect of substitution of Rh with Ir on elastic anisotropy is significant. Based on all indices ScIrP is elastically more anisotropic than ScRhP.

Following Ref. [33], it is possible to calculate the Debye temperature $\theta_D$ as follows:

$$\theta_D = \frac{h}{k_B}\left[\left(\frac{3n}{4\pi}\right)\frac{N_A \rho}{M}\right]^{1/3} v_m$$

where $h$ is the Planck's constant, $k_B$ denotes the Boltzmann's constant, $N_A$ refers the Avogadro's number, $\rho$ is the mass density, $M$ stands for the molecular weight and $n$ refers the number of atoms in the molecule. In a polycrystalline solid, the sound wave travels with an average velocity $v_m$, which can be calculated from

$$v_m = \left[\frac{1}{3}\left(\frac{1}{v_l^3} + \frac{2}{v_t^3}\right)\right]^{-1/3}$$

where $v_l$ and $v_t$ refer the longitudinal and transverse sound velocities in a polycrystalline material, respectively. These velocities can be obtained from the polycrystalline shear modulus $G$ and the bulk modulus $B$ as follows:

$$v_l = \left[\frac{3B + 4G}{3\rho}\right]^{1/2}$$

and

$$v_t = \left[\frac{G}{\rho}\right]^{1/2}$$

The calculated mass density $\rho$, sound velocities $v_l$, $v_t$, and $v_m$ and Debye temperature $\theta_D$ are listed in Table 3. The calculated values of $\theta_D$ for ScRhP and ScIrP are 468 and 397 K, respectively. But, the respective values estimated from the coefficient of the lattice heat capacity $\beta$ are found to be 287 and 250 K [10,11]. It should be noted that the Debye temperature obtained from the elastic constants may vary substantially from those extracted from the fit of lattice heat capacity. The fit to the heat capacity data with a singular Debye temperature always introduces some error [34]. Besides, if there is significant anharmonic contribution to the lattice heat capacity, the error in the estimation of Debye temperature becomes more pronounced [34]. This issue definitely requires further attention. In spite of that, the tendency in both results is almost similar. The replacement of Rh with Ir from ScRhP causes a reduction of 15.2% and 12.9% for $\theta_D$ in theoretical and estimated values, respectively. In most cases, a higher Debye temperature corresponds to a higher phonon thermal conductivity as well as phonon frequency. Therefore, the Rh-containing phosphide should be thermally more conductive than the Ir-containing phosphide and the phonon frequency in ScRhP is expected to be higher than in ScIrP.

**Table 3.** Calculated density ($\rho$ in gm/cm$^3$), longitudinal, transverse and average sound velocities ($v_l$, $v_t$, and $v_m$ in km/s) and Debye temperature ($\theta_D$ in K) of ScRhP and ScIrP.

| Compounds | $\rho$ | $v_l$ | $v_t$ | $v_m$ | $\theta_D$ |
|---|---|---|---|---|---|
| ScRhP | 6.47 | 6.488 | 3.427 | 3.831 | 459, 287[a] |
| ScIrP | 9.80 | 5.527 | 2.893 | 3.236 | 389, 250[a] |

[a]Ref. [11]

### 3.3. *Electronic properties*

The calculated electronic band structures along the high-symmetry directions of the Brillouin zone are depicted in Figs. 2a and b, which exhibit the electronic energy dispersions of ScRhP and ScIrP pnictides, respectively. The Fermi level, $E_F$, is chosen to be at zero of the energy scale. The valence and conduction bands overlap considerably and as a result no band gap is found at the Fermi level. Therefore, both the compounds under study should exhibit metallic conductivity. An appreciably large DOS at $E_F$ in ScRhP compared to that in ScIrP is expected from the fact that the nearly flat bands along L-H direction in ScIrP found to be close to the Fermi level in ScRhP. The overall band profiles for both the new silicide superconductors are almost similar to those found in literature [10,11].

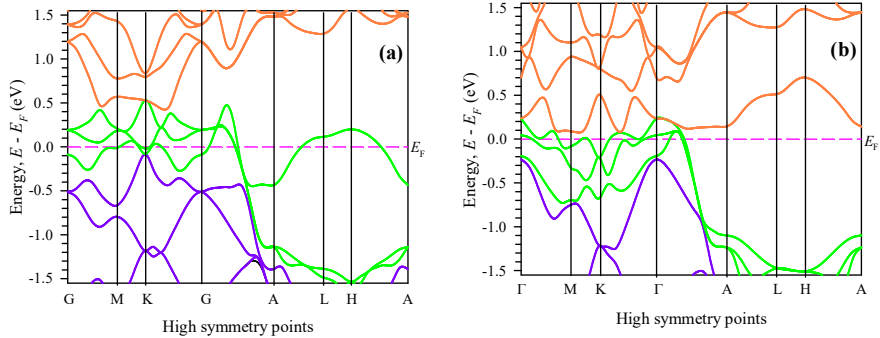

Fig. 2. Electronic band structure of (a) ScRhP and (b) ScIrP.

To further elucidate the nature of chemical bonding in ScRhP and ScIrP, we have calculated the total and partial electronic density of states (DOS) and shown those in Figs. 3a and b. The sharp peak of the DOS at the Fermi level $E_F$ is an indication of structural instability, whereas a deep valley of DOS at $E_F$ implies the structural stability. Therefore, between two pnictides, ScIrP is expected to be more stable structurally. The lowest lying valence band situated between –12.7 to –10.5 eV in ScRhP and between –13.4 to –11.0 eV in ScIrP arises due to P 2$s$ electrons. Then a band gap of width 3.5 eV in ScRhP and 2.3 eV in ScIrP is observed. The highest valence band in both superconductors consists of several distinct peaks. This wide valence band extended to the Fermi level arises due to mutual interaction among P 2*p*, Rh/Ir 4*d*/5*d* and Sc 3*d* states in both ternary compounds. These interactions indicate strong covalent P-Rh and P-Sc bonding in ScRhP and P-Ir and P-Sc bonding in ScIrP. The DOS at the Fermi level is found to be 6.54 and 4.43 states per eV per unit cell for ScRhP and ScIrP, respectively. The corresponding values reported in the earlier study [10,11] are 9.58 and 5.16 states per eV per unit cell when SOC is considered and 9.61 and 4.99 states per eV per unit cell when SOC is ignored. It is seen that all values for ScIrP are almost similar. But a large difference is observed between present and previous results for ScRhP though the patterns of DOS around the Fermi level are almost identical. It is also observed that the inclusion of SOC gives rise an increase in DOS at $E_F$ in ScIrP but a decrease in ScRhP. This oppositeness and discrepancy between present and previous results for DOS at $E_F$ in ScRhP demand more theoretical work.

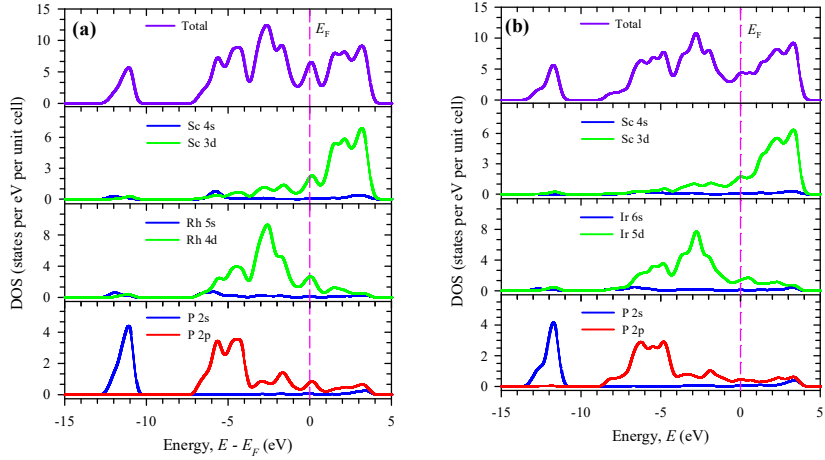

Fig. 3 Total and partial electronic density of states (DOS) of (a) ScRhP and (b) ScIrP.

3.4. *Mulliken populations*

Mulliken population analysis is extensively used to study the bonding characteristics. This method assigns charges to wave functions presented in linear combination of atomic orbitals (LCAO) basis sets. In CASTEP code, it is employed with a projection of the plane wave states onto a localized basis via a technique introduced by Sanchez-Portal *et al*. [35]. The population analysis of the resulting projected states is then carried out with the formalism developed in Mulliken scheme [36]. This analysis provides us with the bond overlap population and the effective valence charge as a gauge of ionicity/covalency of chemical bonds. The Mulliken charge assigned with a certain atomic species $\alpha$ can be evaluated as:

$$Q(\alpha) = \sum_{k} w_k \sum_{\mu}^{on\,\alpha} \sum_{\nu} P_{\mu\nu}(k) S_{\mu\nu}(k)$$

and the bond population involving two atoms $\alpha$ and $\beta$ is

$$P(\alpha\beta) = \sum_{k} w_k \sum_{\mu}^{on\,\alpha} \sum_{\nu}^{on\,\beta} 2 P_{\mu\nu}(k) S_{\mu\nu}(k)$$

where $P_{\mu\nu}$ denotes the density matrix elements and $S_{\mu\nu}$ refers the overlap matrix. The Mulliken charge leads to calculate the effective valence charge as a difference of the formal ionic charge from the Mulliken charge on the anion species. The strength of a chemical bond with ionicity or covalency can be determined with the effective valence. A pure ionic bond exists if the effective valence attains a zero value. A non-zero effective valence is found to be involved with a covalent bond. The degree of covalency of a chemical bond can be assessed with a positive value of effective valence being how much greater than zero. The calculated effective valence for atoms in two new ternary superconducting pnictides is listed in Table 4, which provides an indication of significant covalency in atomic bonding inside two compounds. The Mulliken atomic populations also quantify the charge transfer from one atom to another. As an example, the charge transfer from Sc to Rh is 0.51e, while 0.09e and 0.28e charges are transferred to the two P atoms of two different atomic sites in ScRhP.

Table 4. Mulliken population analysis of ScRhP and ScIrP

| Compounds | Species | Mulliken atomic populations | | | | | Effective valence charge (e) |
|---|---|---|---|---|---|---|---|
| | | $s$ | $p$ | $d$ | Total | Charge (e) | |
| ScRhP | Sc | 2.38 | 6.23 | 1.66 | 10.27 | 0.73 | 1.27 |
| | Rh | 0.83 | 0.39 | 8.30 | 9.51 | −0.51 | 2.49 |
| | P1 | 1.63 | 3.46 | 0.00 | 5.09 | −0.09 | – – |
| | P2 | 1.62 | 3.66 | 0.00 | 5.28 | −0.28 | – – |
| ScIrP | Sc | 2.19 | 6.27 | 1.72 | 10.49 | 0.51 | 2.49 |
| | Ir | 0.67 | 0.84 | 7.93 | 9.43 | −0.43 | 2.57 |
| | P1 | 1.56 | 3.42 | 0.00 | 4.98 | 0.02 | – – |
| | P2 | 1.55 | 3.59 | 0.00 | 5.13 | −0.13 | – – |

The bond overlap population can predict the nature of chemical bonding in crystals. With a negligible value, close to zero, a bond population indicates an insignificant interaction between the electronic populations of two atoms. In fact, a chemical bond of small population is really weak and plays no role in materials' hardness. An overlap population with small value is an indication of ionic bonding. A high level of covalency is associated with a chemical bond when the overlap population carries a high value. The positive and negative overlap populations are responsible for bonding and antibonding states, respectively. The calculated bond overlap populations are given in Table 5. It is observed that the covalent P1-Rh bond is weaker in ScRhP than the similar P1-Ir bond in ScIrP. But, the P2-Rh bond in ScRhP exhibits more covalency than the similar P2-Ir bond in ScIrP. The P2-Sc bond in ScRhP is more covalent than that in ScIrP. In ScRhP, the Rh-Rh bond arises with a reasonable negative population, whereas the similar Ir-Ir bond in ScIrP appears with a small positive population. It means that Rh-Rh is an ionic bond and Ir-Ir is a covalent bond. The other bonds indicate the more ionicity in ScRhP compared to ScIrP. Therefore, we may conclude that the covalent bond dominates in ScIrP compared to ScRhP. This dominating of covalent bonding should make ScIrP as a comparatively hard material.

**Table 5.** Calculated Mulliken bond number $n^\mu$, bond length $d^\mu$, and bond overlap population $P^\mu$ of $\mu$-type bond for ScRhP and ScIrP (with their metallic populations in parenthesis).

| ScRhP ($P^{\mu'}$ = 0.0238) | | | | ScIrP ($P^{\mu'}$ = 0.0184) | | | |
|---|---|---|---|---|---|---|---|
| Bond | $n^\mu$ | $d^\mu$ (Å) | $P^\mu$ | Bond | $n^\mu$ | $d^\mu$ (Å) | $P^\mu$ |
| P1–Rh | 6 | 2.47131 | 0.26 | P1–Ir | 6 | 2.41668 | 0.42 |
| P2–Rh | 3 | 2.49676 | 0.83 | P2–Ir | 3 | 2.52359 | 0.78 |
| P1–Sc | 3 | 2.70150 | −0.08 | P1–Sc | 3 | 2.66032 | −0.05 |
| P2–Sc | 6 | 2.71691 | 0.45 | P2–Sc | 6 | 2.72802 | 0.32 |
| Rh–Rh | 1 | 2.81559 | −0.38 | Ir–Ir | 1 | 2.75967 | 0.09 |
| Sc–Rh | 3 | 2.86868 | −1.78 | Sc–Ir | 3 | 2.86189 | −1.17 |

To predict the level of the metallicity of a chemical bond, we have used the relation, $f_m = P^{\mu'}/P^{\mu}$ [37,38]. In ScRhP, the bonds P1–Rh, P2–Rh and P2–Sc possess the metallicity of 0.0915, 0.0287, and 0.0529, respectively, suggesting that P1-Rh is more metallic than other bonds. The bonds P1–Ir, P2–Ir, P2–Sc, and Ir–Ir in ScIrP exhibit metallicity with 0.0438, 0.0236, 0.0575, and 0.2044 values, respectively. Among these bonds, Ir–Ir has highest metallicity in ScIrP. Based on above discussion, we can come into decision that the new ternary phosphides are characterized as metallic and covalent materials with some ionic nature.

*3.5. Theoretical Vickers hardness*

The hardness of a material is perceptibly a macroscopic concept, which is characterized by both the intrinsic and extrinsic properties. The intrinsic properties include bond strength, cohesive energy and crystal structure. Conversely, the extrinsic properties comprise defects, stress fields and morphology. The experimental values of hardness depend on the methods applied for measurement, temperature, etc. Similarly, the theoretical values are influenced by the formalism used for calculations. For partial metallic compounds like studied ternaries, Gou *et al*. [37] reformulated a formula proposed by Gao [39] for non-metallic covalent materials. This reformulated theory has become popular. According to this method, the bond hardness can be calculated as:

$$H_v^{\mu} = 740\left(P^{\mu} - P^{\mu'}\right)(v_b^{\mu})^{-5/3}$$

In this formula, $P^{\mu}$ is the Mulliken overlap population of the $\mu$-type bond, $P^{\mu'}$ refers the metallic population and is determined with the unit cell volume $V$ and the number of free electrons in a cell $n_{free} = \int_{E_P}^{E_F} N(E)dE$ as $P^{\mu'} = n_{free}/V$, $E_P$ represents the energy at pseudogap, and $v_b^{\mu}$ is the volume of a bond of $\mu$-type, which is estimated using the bond length $d^{\mu}$ of type $\mu$ and the number of bonds $N_b^{\nu}$ of type $\nu$ per unit volume via the equation $v_b^{\mu} = (d^{\mu})^3/\sum_{\nu}[(d^{\mu})^3 N_b^{\nu}]$. If a crystal consists of the complex multiband then its hardness can be obtained as a geometric average of harnesses for all bonds as follows [40,41]:

$$H_V = [\prod^{\mu}(H_v^{\mu})^{n^{\mu}}]^{1/\Sigma n^{\mu}}$$

where $n^{\mu}$ is the number of $\mu$-type bonds, which compose the crystals of multiband. Along with the individual bond hardness, the hardness value of two superconducting pnictides is listed in Table 6. These results include only the positive as well as reasonable populations between the nearest neighbors (atoms) in the first coordination shells. The calculated hardness values for ScRhP and ScIrP are 7.13 and 7.78 GPa, respectively, indicating that the Ir-based phosphide is harder than the Rh-based phosphide, which has already been predicted from shear modulus and chemical bond analysis.

Table 6. Calculated bond and Vickers hardness $H_v^\mu$, $H_v$ (in GPa) of ScRhP and ScIrP along with bond number $n^\mu$, bond length $d^\mu$ (Å), bond volume $v_b^\mu$ (Å³) and bond as well as metallic populations $P^\mu$, $P^{\mu'}$.

| Compounds | Bond | $n^\mu$ | $d^\mu$ (Å) | $P^\mu$ | $P^{\mu'}$ | $v_b^\mu$ (Å³) | $H_v^\mu$ (GPa) | $H_v$ (GPa) |
|---|---|---|---|---|---|---|---|---|
| ScRhP | P–Rh | 6 | 2.47131 | 0.26 | 0.0238 | 8.07 | 5.38 | 7.13, 6.76[a] |
|  | P–Rh | 3 | 2.49676 | 0.83 | 0.0238 | 8.32 | 17.46 |  |
|  | P–Sc | 6 | 2.71691 | 0.45 | 0.0238 | 10.72 | 6.05 |  |
| ScIrP | P–Ir | 6 | 2.41668 | 0.42 | 0.0184 | 7.55 | 10.23 | 7.78, 6.85[a] |
|  | P–Ir | 3 | 2.52359 | 0.78 | 0.0184 | 8.60 | 15.61 |  |
|  | P–Sc | 6 | 2.72802 | 0.32 | 0.0184 | 10.86 | 4.19 |  |

[a]Calculated according to Chen *et al.* [42].

The existence of antibonding Rh–Rh in ScRhP with Mulliken overlap population of –0.38 may give rise to decrease in the hardness of ScRhP. Therefore, ScRhP is soft and easily mechinable than ScIrP. Additionally, we have calculated the Vickers hardness using the formula developed by Chen *et al.* [42]: $H_V = 2(k^2 G)^{0.585} - 3$ with $k = G/B$, which is recently paying attention of the scientific community. This scheme also gives almost similar results for the two pnictides under study.

*3.6. Charge Density*

The mapping image of electron charge density distribution assists us to predict the nature of chemical bonding in crystals [43]. The contour maps (in the units of e/Å³) of calculated electron charge density distribution for ScRhP and ScIrP have been depicted in Figs. 4 and 5, respectively. To guess the intensity of the charge density for electrons, a colored scale is shown to the adjacent of the contour maps. In this scale, the low and high densities of electronic charge are indicated by the red and blue colors, respectively.

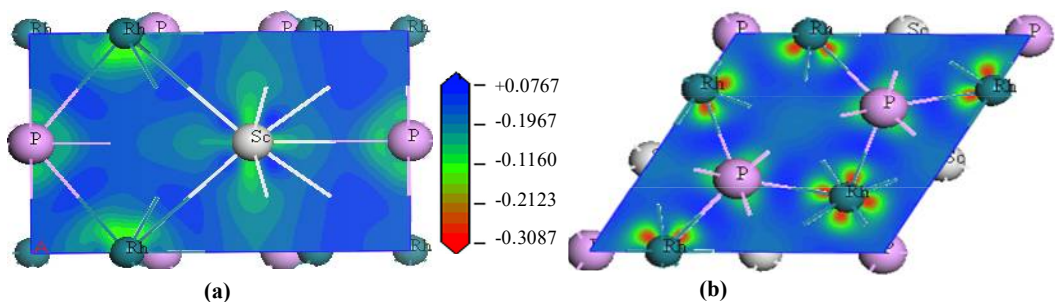

Fig. 4 Electron charge density maps of ScRhP (a) for (100) and (b) for (001) plane.

The strong covalent P–Rh and P–Sc bonding in ScRhP is identified clearly from the accumulation of the charge around the relevant atoms in Figs. 4a and b. The spherical charge distribution around two adjacent Rh atoms (in Fig. 4b) indicates the ionic Rh–Rh bonding in ScRhP. Another ionic bonding between Sc and Rh is expected due to charge balance (undistorted electron clouds) at their positions. However, the P atoms of 2c atomic site are found to form an ionic bond with Sc though the same atom of 1b Wyckoff position makes a covalent bond with Sc, which is also evident from the calculated Mulliken bond population. The similar bonding features are observed for iso-structural ScIrP in Figs. 5.

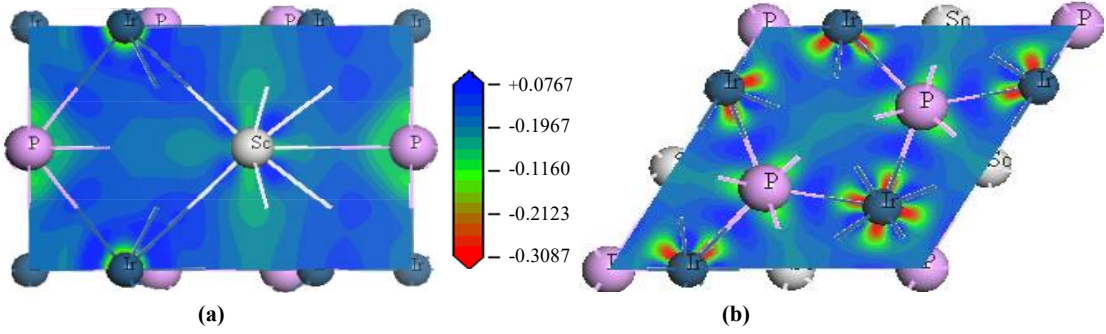

Fig. 5 Electron charge density maps of ScIrP (a) for (100) and (b) for (001) plane.

3.7. *Electron-phonon coupling*

The electron-phonon coupling constant, $\lambda$ is an important parameter for a superconductor. To estimate the electron-phonon coupling constant for the two ternary pnictides, we have used the equation due to McMillan [44] given by,

$$\lambda = \frac{1.04 + \mu^* \left(\frac{\theta_D}{1.45 T_c}\right)}{(1 - 0.62\mu^*)\left(\frac{\theta_D}{1.45 T_c}\right) - 1.04}$$

Here, $\mu^*$ is the repulsive electron-electron Coulomb pseudopotential. We have utilized the experimentally observed $T_c$ values of 2.0 K [11] and 2.95 K (average of the two reported values) [10,21] for ScRhP and ScIrP, respectively, with $\mu^* = 0.10$ and the theoretically calculated Debye temperatures. This procedure yields $\lambda = 0.414$, and 0.466, for ScRhP and ScIrP, respectively. It is worth noticing that $\lambda$ is slightly higher for ScIrP compared to that of ScRhP, even though the Debye temperature and $N(E_F)$ of ScRhP are significantly higher. This implies that the matrix elements of electron-phonon interaction are markedly enhanced in ScIrP compared to that in ScRhP.

## 4. Conclusion

In summary, the first principles calculations have been performed to explore the structural, elastic, electronic, and superconducting properties of the newly synthesized superconducting ScRhP and ScIrP ternary pnictides. Our calculated structural parameters are in good agreement with the experimental results. The calculated elastic constants satisfy the mechanical stability conditions for the hexagonal ScRhP and ScIrP. Based on the calculated bulk modulus, shear modulus, Young's modulus, Poisson's ratio, Pugh's ratio, and elastic anisotropic factors, we have evaluated the mechanical behaviors of these ternaries. Both the superconducting pnictides are ductile in nature and anisotropic elastically. ScRhP is predicted to be thermally more conductive than ScIrP and the phonon frequency in ScRhP is anticipated to be higher than in ScIrP. From the analysis of the elastic constants and moduli and Vickers hardness, we have found that ScRhP is softer than ScIrP, and therefore, is comparatively easily mechinable than ScIrP. The calculated electronic features show that the metallic conductivity of ScRhP decreases considerably when Rh is replaced with Ir. The major contribution to the TDOS at $E_F$ comes from d-electrons of transition metals in both compounds. The calculated DOSs, Mulliken populations, and charge density maps imply that the chemical bonding in the two phosphide superconductors can be described as a mixture of covalent, metallic and ionic in nature. The calculated electron-phonon coupling constants indicate that both these compounds are moderately coupled BCS superconductors. The electron-phonon interaction potential appears to be much stronger in ScIrP compared to ScRhP. It is quite interesting to note that the superconducting state properties of these two ternary phosphides are almost identical to those of recently discovered hexagonal ternary silicide compounds, $Li_2IrSi_3$ and $Li_2PtSi_3$ [45,46]